# Band structure effects in nitrogen *K*-edge resonant inelastic X-ray scattering from GaN

V.N. Strocov[1], T. Schmitt[2], J.-E. Rubensson[2], P. Blaha[3], T. Paskova[4] and P.O. Nilsson[5]

[1] Paul Scherrer Institute, CH-5232 Villigen PSI, Switzerland
[2] Uppsala University, Box 530, SE-75121 Uppsala, Sweden
[3] Institut für Materialchemie, Technische Universität Wien, A-1060 Wien, Austria
[4] Linköping University, SE-581 83 Linköping, Sweden
[5] Chalmers University of Technology, SE-412 96 Göteborg, Sweden

Systematic experimental data on resonant inelastic X-ray scattering (RIXS) in GaN near the N *K*-edge are presented for the first time. Excitation energy dependence of the spectral structures manifests the band structure effects originating from momentum selectivity of the RIXS process. This finding allows obtaining **k**-resolved band structure information for GaN crystals and nanostructures.

**Introduction** Soft-X-ray emission (SXE) and absorption (SXA) spectroscopies allow investigation of the electronic structure in the valence band (VB) and conduction band (CB), respectively, with elemental specificity and large probing depth (1000-3000 Å). In particular, buried nanoclusters, interfaces (see [1] and further references in a review [2]) and even isolated impurities [3] can be accessed by these techniques. In general, they characterize the electron density of states averaged in the wavevector **k**. For wide band gap semiconductor systems, distinct **k**-selectivity appears nevertheless in resonant inelastic X-ray scattering (RIXS) manifested by the SXE spectral dependence on excitation energy within a few eV from the absorption threshold [2]. In RIXS, the absorption and emission events are coupled in a fast coherent process, in which the full momentum is conserved as

$$\mathbf{q}_{in} = \mathbf{q}_{out} + \mathbf{k}_e + \mathbf{k}_h$$

where $\mathbf{q}_{in}$ and $\mathbf{q}_{out}$ are the absorbed and emitted photon wavevectors, respectively, and $\mathbf{k}_e$ and $\mathbf{k}_h$ are the electron and hole wavevectors. The **k**-selectivity of RIXS has been employed to study band structures of a number of semiconductors and semimetal materials such as Si, SiC and graphite [2,4,5].

GaN is a semiconductor with a wide range of device applications such as the blue light emitting diodes. However, there is still a lack of detailed knowledge on the basic electronic properties of nitride compounds. Studies of the band structure of GaN with resolution in the **k**-space using conventional angle-resolved photoelectron spectroscopy are burdened by surface reconstructions. The **k**-selectivity of RIXS, surprisingly, has so far not been exploited in any systematic study on GaN, although a pilot study [6] has given an evidence of such phenomena.

**Experiment** Our sample was a 30 μm thick wurtzite GaN film grown by hydride vapour phase epitaxy on sapphire substrate. The SXE/SXA experimental data were taken near the N *K*-edge (1*s* core level) at ~400 eV. The experiment was performed in MAX-lab, Sweden, at the beamline I511-3 equipped with a high-resolution Rowland-mount SXE spectrometer [7] installed in the plane of incidence at a scattering angle of 90°. The monochromator and spectrometer were both operated with a resolution around 0.3eV, and were calibrated in absolute energy scale as described in [3]. The incident light was *p*-polarized and made a grazing angle of 20° to the sample surface.

The experimental SXA spectrum is shown in Fig.1 (left panel). With the N 1*s* core level, it represents the N *p*-projected partial density of states (PDOS) in the CB. Moreover, in highly anisotropic wurtzite GaN the SXA structures depend on the relative weight of the $p_z$ and $p_{xy}$ orbital contributions, varying with the incidence angle [8]. Our SXA spectrum, measured near the **E**||**c** condition, has the largest contribution from the $p_z$ orbitals and thus reflects predominantly the N $p_z$ PDOS. The experimental off-resonant SXE spectrum, taken with an excitation energy $h\nu_{ex}$ of 420 eV, is shown in Fig.1 (left). Large phase space for *e-ph* and *e-e* interactions in the SXE intermediate state, available with such $h\nu_{ex}$ well above the absorption threshold, results in vanishing of the coherent emission fraction by effective averaging over the **k**-space. Our SXE spectrum is therefore incoherent, representing the N *p* PDOS in the VB. Moreover, measured near the **E**⊥**c** condition, it reflects predominantly the N $p_{xy}$ PDOS. Our SXA and off-resonant SXE data are in good correspondence with the previous experiments [9,10].

Our experimental resonant SXE series, taken with $h\nu_{ex}$ in a few eV above the absorption threshold, is shown on top of the off-resonant SXE spectrum. Clear dispersions and lineshape changes of both spectral peaks with $h\nu_{ex}$ identify pronounced RIXS phenomena.

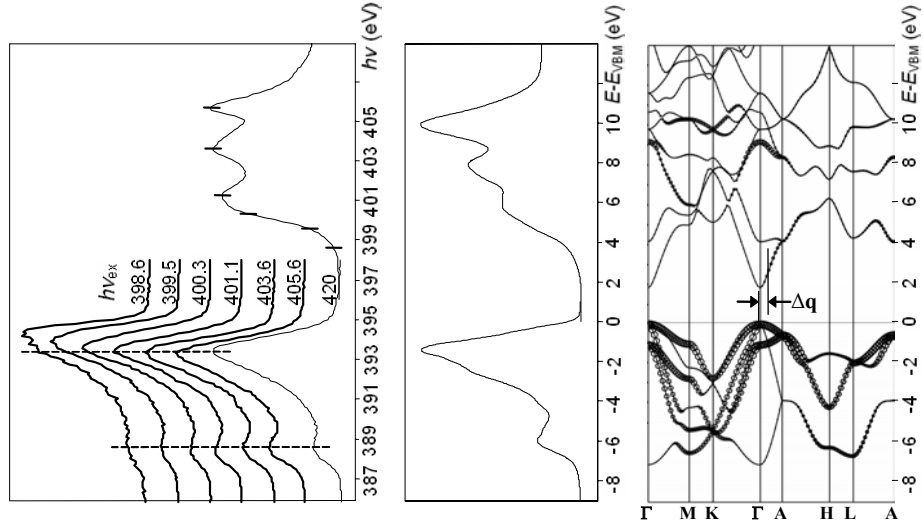

Fig.1: (*left panel*) Experimental SXA and off-resonant SXE spectra (thin lines), and the resonant SXE series (bold) at the indicated excitation energies (also marked by ticks on the SXA spectrum). The dashed lines show the off-resonant energies of the SXE peaks; (*middle*) Calculated $p_z$ SXA and off-resonant $p_{xy}$ SXE spectra, relevant with our experimental geometry; (*right*) Calculated band structure, with the N $p_z$ weights of the CB states and N $p_{xy}$ of the VB states represented by radii of the circles.

**Calculations** The calculations were performed within the GGA-DFT framework using the full-potential LAPW+local orbitals method implemented in the WIEN2k package [11]. Apart from the band structure, we have calculated the SXA spectrum assuming filled core hole, and the off-resonant SXE spectrum assuming the 'final-state rule' with filled core hole and screened valence hole. In this framework, the spectra are essentially the N $p$-PDOS multiplied by smoothly varying matrix elements. In view of our experimental geometry, the SXA calculation included only the $p_z$ orbitals, and SXE only the $p_{xy}$ ones. To simulate the lifetime broadening, the calculated spectra were smoothed by Lorentzian with fullwidth, taken to linearly increase with energy distance from the Fermi level, adjusted to the experimental spectral broadening. The experimental resolution was simulated by Gaussian smoothing.

The calculated $p_z$ SXA and $p_{xy}$ SXE spectra, aligned in energy with the experiment at the leading edge of the SXE spectrum, are shown in in Fig.1 (middle panel). They are in convincing agreement with the experiment. Systematic shifts of the theoretical SXA structures to lower energies are along with the well-known band gap problem of the DFT, manifesting the excited-state self-energy corrections to the static DFT exchange-correlation. In both VB and CB such corrections show some energy dependence.

The calculated band structure $E(\mathbf{k})$ is shown in Fig.1 (right). The N $p_z$ and N $p_{xy}$ weights of the CB and VB states respectively, relevant with our experimental geometry, are indicated by circles.

**k-selectivity effects in RIXS** Analysis of the band structure effects for GaN is somewhat complicated by that each SXE and SXA spectral peak piles up due to a multitude of critical points with high DOS along different Brillouin zone directions, and can not be associated with one certain **k**-point. Nevertheless, the resonant series in Fig.1 unambiguously identifies **k**-selectivity effects:

(1) most importantly, tuning $h\nu_{ex}$ on the absorption onset at 398.6 eV results in shifting of the dominant SXE peak to its highest energy. As seen from comparison of the calculated SXA spectrum to the N $p_z$ band structure, for this $h\nu_{ex}$ the excited electrons appear in the CB already in some 1 eV above the CB minimum (such a delayed onset occurs because the lowest CB states are depleted in the N $p$ character) with the corresponding **k**-vectors distributing near ΓA in some 0.3 Å$^{-1}$ from the Γ-point. The photon wavevector transfer $\Delta\mathbf{q}=\mathbf{q}_{out}-\mathbf{q}_{in}$ in our experiment has almost the same value 0.28 Å$^{-1}$, and is also directed close to ΓA. Therefore, as indicated in Fig.1, the **k**-conserving RIXS process couples the excited N $p_z$ conduction electrons to the N $p_{xy}$ valence holes in the VB maximum in the Γ-point, which results in a strong coherent emission from the VB maximum appearing on the high-energy side of the main incoherent peak. In the experimental resonant series this is seen as shifting of the SXE spectral maximum to its highest energy position. The VB minimum, placed in the same Γ-point but having the $p_z$ character, shows only a very slight coherent intensity pile up due to deviation from the ideal **E**⊥**c** geometry;

(2) with increase of $h\nu_{ex}$ the main SXE peak disperses to lower energies, reflecting the valence band dispersions with **k** moving away from the Γ-point along ΓAH. Its dispersion range towards $h\nu_{ex}$ of 401eV is about 0.9 eV, which, with the calculated CB dispersions, well matches the dispersion of the $p_{xy}$ heavy-hole valence bands (possibly renormalized by coupling to the core hole). The light-hole valence band is not seen in the SXE spectra because of its $p_z$ character in this region of the **k**-space;

(3) with $hv_{ex}$ varied near 401 eV, the SXE spectra display a sizeable dispersing coherent fraction below the low-energy incoherent peak. Based on the N $p_z$ dispersions in the CB, it can be associated with the N $p_{xy}$ dispersions in the M-valley near the VB bottom.

Interestingly, differences of the resonant SXE spectra from the off-resonance one, indicating the coherent fraction, remain notable as far as up to 7 eV from the absorption threshold.

**Conclusion** Systematic high-resolution SXE and SXA experimental data on wurtzite GaN near the N *K*-edge are presented for the first time. The resonant SXE spectral structures show up pronounced dispersions and lineshape changes with the excitation energy, identifying **k**-selectivity effects in RIXS. Our finding allows **k**-resolved band structure studies of GaN with implications for its optoelectronic properties. With the elemental specificity and large probing depth of the soft-X-ray spectroscopies, it can also be used, for example, on buried GaN nanoclusters to control their direct respective indirect band gap.